\documentclass[twocolumn]{aa}                  

\usepackage{multirow}
\usepackage{graphicx}
\usepackage{rotating}
\usepackage{subfigure}
\usepackage[flushleft]{threeparttable}
\usepackage{color}
\usepackage{txfonts}
\usepackage[square,comma,numbers,sort&compress]{natbib}
\usepackage{epstopdf}

\bibpunct{(}{)}{;}{a}{}{,}                      
\begin{document}
   \title{Dielectronic recombination data for dynamic finite-density plasmas}
    \subtitle{XV. The silicon isoelectronic sequence}

   \author{Jagjit Kaur$^1$, T.~W.~Gorczyca$^1$, and N.~R.~Badnell$^2$}

   \institute{
   $^1$Department of Physics, Western Michigan University, Kalamazoo, MI 49008, USA\\
   $^2$Department of Physics, University of Strathclyde, Glasgow G4 0NG, UK\\
   \email{gorczyca@wmich.edu}
   }

   \date{}


  \abstract
   {
   We aim to present a comprehensive theoretical investigation of dielectronic recombination (DR) of the silicon-like 
isoelectronic sequence and provide DR and radiative recombination (RR) data that can be used within a 
generalized collisional-radiative modelling framework. 
   }
   {
   Total and final-state level-resolved DR and RR rate coefficients for the ground and metastable initial levels 
of 16 ions between $\rm{P^{+}}$ and $\rm{Zn^{16+}}$ are determined.
   }
   {
   We carried out multi-configurational Breit-Pauli (MCBP) DR calculations for silicon-like ions in the 
independent processes, isolated resonance, distorted wave (IPIRDW) approximation. Both $\Delta n_c =0$ and 
$\Delta n_c =1$ core excitations are included using $LS$ and intermediate coupling  schemes.
   }
   {Results are presented for a selected number of ions and compared to all other existing theoretical and 
experimental data. The total dielectronic and radiative recombination rate coefficients for the ground state 
are presented in tabulated form for easy implementation into spectral modelling codes. These data can also 
be accessed from the Atomic Data and Analysis Structure (ADAS) OPEN-ADAS database. This work is a part of an assembly of 
a dielectronic recombination database for the modelling of dynamic finite-density plasmas.   
   }
   {}

   \authorrunning{J. Kaur et al.}
   \maketitle

\section{Introduction}
\label{Sec:Intro}

The emission of electromagnetic radiation from cosmic plasmas, as a consequence of electron-ion collision 
processes, reveals information about physical parameters of the plasma such as chemical composition, pressure, 
electronic or ionic temperature, and density. However, the accuracy of these parameters is strongly 
influenced by uncertainties in the ionization balance calculations, which are in turn affected by 
uncertainties in the ionization and recombination rate coefficients. Therefore, it is of fundamental 
interest to determine accurate rate coefficients for astrophysical and laboratory plasma studies.

Dielectronic recombination (DR) --- \cite{Burgess:1964, Burgess:1965} --- is the dominant electron-ion 
recombination process in most photoionized and (electron) collisionally ionized plasmas. In collisionally 
ionized plasmas 
(e.g. stars and supernovae remnants), 
the ionization 
occurs due to electrons and ions that are formed at a temperature approximately half of their ionization 
potential \citep{Bryans:2006}. On the other hand, in photoionized plasmas (e.g. \ion{H}{ii} regions and 
planetary nebulae), ionization occurs due to photons and ions that are formed at temperatures below the 
ionization energies \citep{CLOUDY:90, Kallman:2001}. Therefore, the ionization balance is achieved over 
very different temperature ranges in collisionally ionized and photoionized plasmas. To model the spectral 
emission, plasma modelling codes, including CLOUDY \citep{CLOUDY:90} and XSTAR \citep{Kallman:2001}, for 
photoionized plasmas, and the CHIANTI code \citep{Landi:2006}, for collisionally ionized plasmas, require 
accurate DR rate coefficients over a wide range of temperatures to determine the elemental abundances and,
therefore, plasma parameters. 

In order to address the need for accurate DR rate coefficients, a large collaborative effort was initiated by 
\cite{Badnell:2003} to calculate the total and final-state level-resolved DR rate coefficients from the ground 
and the metastable states of all ionization stages of all ions up to \ion{Zn}{} relevant to the modelling of 
astrophysical and laboratory plasmas. In a series of papers, multiconfiguration Breit-Pauli 
(MCBP) calculations have been performed to provide a DR database for all isoelectronic sequences of the first 
and second row ions and third row ions up through \ion{Al}-like \citep{Badnell:H:2006, Bautista:He:2007, 
Colgan:Li:2004, Colgan:Be:2003, Altun:B:2004, Zatsarinny:C:2004, Mitnik:N:2004, Zatsarinny:O:2003, 
Zatsarinny:F:2006, Zatsarinny:Ne:2004, Altun:Na:2006, Altun:Mg:2007, Shahin:2012} and also \ion{Ar}-like 
\citep{Dragon:Ar}. Final-state level-resolved 
DR rate coefficients are necessary for modelling plasmas within the collisional-radiative framework at densities 
found in astrophysical plasmas, such as solar flares ($>10^{12}~\rm{cm}^{-3}$) \citep{Polito:2016a, Polito:2016b}, 
and in magnetic fusion plasmas for example ITER ($10^{12} - 10^{15}~\rm{cm}^{-3}$) \citep{Watts:2013}, where the coronal 
approximation is not valid. DR data for initial metastable states are required for modelling plasmas with timescales 
comparable to the life time of the metastable states. The first unaddressed isoelectronic sequence in the third 
row is silicon-like, for which no systematic calculations have been performed. Here, we present a reliable DR 
database for the Si-like isoelectronic sequence.

\cite{Bryans:2009a} have demonstrated the effect of inaccuracies in DR data for singly-charged ions in the 
low-temperature regime of molecular clouds, causing significant differences in the abundances of species found 
on the surface of dust grains and in the gas-phase. They have used RR rate coefficients of singly-charged ions 
from the UMIST database \citep{UMIST:2007} in chemical models; however, the origin of this RR data is unclear. 
There are no other RR data, while DR data exist for {\bf $\rm{P^{+}}$}, {\bf $\rm{S^{+}}$}, {\bf $\rm{Cl^{+}}$}, and {\bf $\rm{Fe^{+}}$}. 
Our present study of the silicon-like isoelectronic sequence finally provides state-of-the-art calculations of 
the RR and DR rate coefficients for {\bf $\rm{P^{+}}$}, for instance, that are needed to constrain the chemical models 
used to study the evolution of dense molecular clouds, protostars, and diffuse molecular clouds. Such studies 
are in turn important for understanding the origin of the first organic molecules. 

Furthermore, the difference in the sulphur abundances in planetary nebulae derived from an ionization correction 
factor (ICFs), and the spectral emission line measurements from the expected value observed by \cite{Henry:2012}, 
constitute the motivation to update the {\bf $\rm{S^{2+}}$} DR data, as was done for totals by \cite{Badnell:2015}. \cite{Henry:2012}
demonstrated how uncertainties in the positions of low-lying resonances affect the low-temperature DR rate coefficients, 
that in turn affect the elemental abundances in planetary nebulae. The present study extends the previous 
theoretical work by determining partial as well the total rate coefficients, and including the $\Delta n_c =1$ core excitation, in 
addition to  $\Delta n_c =0$ core excitation.

Additionally, accurate DR rate coefficients for M-shell {\bf $\rm{Fe^{12+}}$} are needed to accurately model the absorption 
features needed to reproduce the so-called unresolved transition array (UTA). This is a series of inner-shell absorption 
lines at $15-17$ {\AA}, caused by $2p\rightarrow 3d $ photoabsorption in the X-ray spectrum of Active Galactic 
Nuclei (AGN), as observed by Chandra and XMM-Newton. The problem has been attributed in 
part to the underestimated low-temperature DR rate coefficients for M-shell \ion{Fe}{} used in the photoionization 
models \citep{Badnell:Fe3pq:2006}. The recent benchmark theoretical and experimental total DR results are presented by \cite{Hahn:2014} for M-shell 
{\bf $\rm{Fe^{12+}}$} ions.

The remainder of this paper is organized as follows: In Sect.~\ref{Sec:theory} we discuss the theoretical 
methodology and outline the present calculations. We then present the results for total dielectronic and radiative 
recombination rate coefficients and compare with earlier theoretical and experimental results in Sect.~\ref{Sec:Analysis}. 
Finally, we summarize the assembly of final data in Sect.~\ref{Sec:Summary}.

\section{Theory}
\label{Sec:theory}
A detailed description of our theoretical calculations has already been given by \cite{Badnell:2003}. 
Here we outline only the main points. The atomic structure and collision code AUTOSTRUCTURE \citep{Badnell:2011} 
was used to perform DR calculations. A multi-configuration Breit-Pauli (MCBP) method is implemented within 
an independent processes, isolated resonance, distorted-wave (IPIRDW) approximation, whereby radiative and 
dielectronic recombination processes are treated independently, neglecting interference between the two, 
which is valid for plasma applications \citep{Pindzola:1992}. The code is based on lowest-order perturbation 
theory, for which both the electron-photon and electron-electron interactions are treated to first order. 
Energy levels, radiative rates, and autoionization rates were calculated in $LS$ and  intermediate coupling (IC) 
approximations. We note that the spin-independent mass-velocity and Darwin relativistic operators are included
 in $LS$ coupling, as well as in IC coupling.  The wave functions for the $N$-electron target system are written as a 
configuration expansion,
\begin{eqnarray}
\Psi_i= \sum\limits_{j=1}^N c_{ij} \phi_j \quad ,
\end{eqnarray}
where $c_{ij}$ are the mixing coefficients that are chosen so as to diagonalize $\langle\Psi_i\vert H \vert\Psi_j~\rangle$, 
where $H$ is the Breit-Pauli Hamiltonian. The set of basis functions are constructed from Slater determinants using 
the one-electron spin-orbitals. 

The dielectronic recombination process for silicon-like ions can be represented schematically as
\begin{eqnarray}
{\rm{e}}^{-} + {\rm{X}}^{z+}(3s^{2}3p^{2}\ ^{3\!}P_{J}) \rightleftharpoons  {\rm{X}}^{(z-1)+\ast\ast} 
\rightarrow {\rm{X}}^{(z-1)+} + h\nu\ ,
\end{eqnarray}
where $z$ represents the degree of ionization for the ion $X$. The basis set consisting of the 
$3s^{2}3p^{2}$, 
$3s^{2}3p3d$, 
$3s3p^{3}$, 
$3s3p^{2}3d$, 
$3s^{2}3d^{2}$, 
$3s3p3d^{2}$, 
$3p^{4}$ and 
$3p^{3}3d$
configurations (assuming a closed shell \ion{Ne}-like core) was used to define the silicon-like target states,
for the (by far dominant) $\Delta n_c =0$ core excitation. 
The one-electron spin-orbitals were obtained using the Thomas-Fermi-Dirac-Amaldi (TFDA) model potential 
\citep{Eissner:1969}, and were optimized by varying the scaling parameters $\lambda_{nl}$ so as to reproduce 
the fine-structure splitting of the $3s^{2}3p^{2}\,^3P_J$ levels, to within 0.0005 Ryd compared to NIST. 
Table \ref{table:lamda} lists the optimized scaling parameters for the entire isoelectronic sequence. 

The $(N+1)$-electron basis was constructed by coupling a valence orbital, $nl$,
or a continuum orbital, $\epsilon l$, to the $N$-electron target configurations, and also included the 
$3s^{2}3p^{3}$, 
$3s^{2}3p^{2}3d$,
$3s^{2}3p3d^{2}$, 
$3s3p^{4}$, 
$3s3p^{3}3d$, 
$3s3p^{2}3d^{2}$, 
$3s^{2}3d^{3}$, 
$3s3p3d^{3}$,
$3p^{5}$,
$3p^{4}3d$ and
$3p^{3}3d^{2}$
configurations. Distorted wave calculations were performed to generate 
the bound $nl$ ($n>3$) and continuum orbitals. The wave functions constructed using this $(N+1)$-electron basis
were used to determine the autoionization and radiative rates, which are then assembled to obtain the final-state 
level-resolved and total dielectronic recombination rate coefficients for all silicon-like ions.

For the valence electron, $n$-values were included up to $25$, and a quantum defect approximation for 
high $n$ up to $1000$ was used \citep{Badnell:2003}. The values for the orbital quantum numbers were included 
up to $l=8$. For intershell ($\Delta n_c=1$) core excitation from the $n=3$ shell, the $N$-electron target 
basis set was comprised of $3s^{2}3p^{2}$, $3s^{2}3p3d$, $3s3p^{3}$, and $3s3p^{2}3d$ configurations in addition 
to configurations arising from $3\ell\rightarrow 4\ell'$ excitations (for $\ell=0-1$ \& $\ell'=0-3$). 
The $(N+1)$-electron target basis was described by coupling a valance orbital $4\ell'$ to the $N$-electron 
configurations for ($\Delta n_c=1$) core excitation plus either coupling a valence orbital $nl$, or a continuum 
orbital $\epsilon l$, to the $N$-electron target configurations. Values of the principal quantum number
included were $n \leq 25$, and  of the continuum/valence electron orbital angular momentum were 
$ \ell' \leq 5$. A quantum defect approximation is included for  $ 25 < n \leq 1000$. 

The partial dielectronic recombination rate coefficient $\alpha_{if}$ from an initial state $i$
to a final, recombined state $f$ is given in the IPIRDW approximation as \citep{Burgess:1964}
\begin{eqnarray} \label{Eq:rate} 
\alpha_{if}(T)& = & \left(\frac{4\pi a_{0}^{2}I_H}{k_B T}\right)^{3/2}\,\sum_{d}\frac{\omega_d}{2\omega_i} 
\exp\left(-\frac{E_c}{k_B T_e}\right) \nonumber \\
& & \times \frac{\sum_{\ell}A^{a}_{d \to i, E_{c \ell}}\,A^{r}_{d \to f}}{\sum_{h} A^{r}_{d \to h} + 
\sum_{m,\ell} A^{a}_{d \to m,E_{c \ell}}}\;,
\end{eqnarray}
where the outer sum is over all accessible $(N+1)$-electron doubly excited resonance states $d$, of statistical weight $\omega_d$, 
$\omega_i$ is the statistical weight of the $N$-electron target state, $A^a$ and $A^r$ are the autoionization and 
radiative rates (the sums over $h$ and $m$ gives rise to the total widths), and $E_c$ is the energy of the 
continuum electron, which is fixed by the position of the resonances. Here, $I_H$ is the ionization potential 
energy of the hydrogen atom, $k_B$ is the Boltzmann constant, and $T$ is the electron temperature. The total 
dielectronic recombination rate coefficient is obtained by summing over all the recombined final 
states $f$,
\begin{eqnarray}
\alpha^{(tot)}_{i}(T) = \sum_{f} \alpha_{if}(T)\,.
\end{eqnarray}

Partial and total RR rate coefficients were also computed using the same $N$- and $(N+1)$-electron configurations 
as for the  $\Delta n_c=0$ core excitation DR calculations, but with no doubly-excited (resonance) states 
${\rm{X}}^{(z-1)+\ast\ast}$.

\section{Results}
\label{Sec:Analysis}
The final-state level-resolved partial dielectronic recombination rate coefficients, from both ground and 
metastable initial levels, were computed and then tabulated in the ADAS \citep{Summers:2003} \emph{adf09} format. 
The total ($\Delta n_c=0$ plus $\Delta n_c=1$) intermediate coupling DR rate coefficients were also fitted 
according to the formula
\begin{equation}\label{Eq:Fit:DR}
 \alpha^{{\rm DR}}(T) = \frac{1}{T^{3/2}} \sum_{i} c_{i} \exp\left(- \frac{E_{i}}{T}\right)~, \qquad(i \leq 8)\,~.
\end{equation}
The fitting coefficients $c_i$  and $E_i$ for DR rate coefficients from the ground state 
are listed in Table \ref{table:dr_all} for the entire silicon-like isoelectronic sequence. Our fits reproduce 
the actual computed data to better than $5\%$ for all ions over the temperature range $z^2(10^1-10^7)$ K, 
where $z$ is the residual charge of the recombining ion. In fact, the accuracy is better than $1\%$ over the 
collisionally-ionized plasma region.

Also, the total RR rate coefficients were computed, tabulated in ADAS format, and fitted using the formula of
\cite{Verner:1996},
\begin{equation}\label{Eq:Fit:RR}
 \alpha^{{\rm RR}}(T) = A \, \sqrt{T_0/T} \, \left[ \left( 1 + \sqrt{T/T_{0}} \right)^{1-B}
 \left(1 + \sqrt{T/T_{1}} \right)^{1+B} \right]^{-1} \;,
\end{equation}
where, for low-charge ions, we replace $B$ by \citep{Gu:2003},
\begin{equation}\label{Eq:Fit:RR-low}
 B \rightarrow B + C \exp(-T_2/T).
\end{equation}
Partial RR rate coefficients are tabulated according to the ADAS
\citep{Summers:2003} \emph{adf48} format. The RR fitting coefficients are also listed in Table \ref{table:rr_all}. 
These fits are accurate to better than $5\%$ over the temperature range $z^2(10^1-10^7)$ K.

We compare our present IC total Maxwellian-averaged DR rate coefficients of a selected number of ions along the 
silicon-like isoelectronic sequence to other available theoretical and experimental results. 
In particular, we compare to the widely used recommended data of \cite{Mewe:1980} and \cite{Mazzotta:1998}. 
\cite{Mewe:1980} developed a single fitting 
formula, based on the data of \cite{Ansari:1970} and \cite{Jacobs:1977}, 
for all ions and for all temperatures. 
The previously recommended database of \cite{Mazzotta:1998} was derived from the calculations of \cite{Jacobs:1977,
Jacobs:1979, Jacobs:1980}, which were then fitted by \cite{Shull:1982}, for even numbered nuclei, 
and interpolated to provide the data  for odd numbered nuclei by \cite{Landini:1991}. 
Also, indicated in the figures are the temperature regions 
of collisionally-ionized and photoionized plasmas. 
These temperature ranges are determined for each ion by considering the range of temperatures for which the ion's 
fractional abundance is 90\% or more of its maximum value.
The collisionally-ionized zones were obtained using the calculations of \cite{Bryans:2009}, 
and the photoionized zones have been computed using CLOUDY \citep{CLOUDY:90}. We note that the DR data used in those \ion{Si}-like abundance calculations were those of \citet{Mazzotta:1998}, 
not including our present DR rate coefficients.

\begin{figure*}[!hbtp]
\centering
\includegraphics[angle=0,scale=1.0]{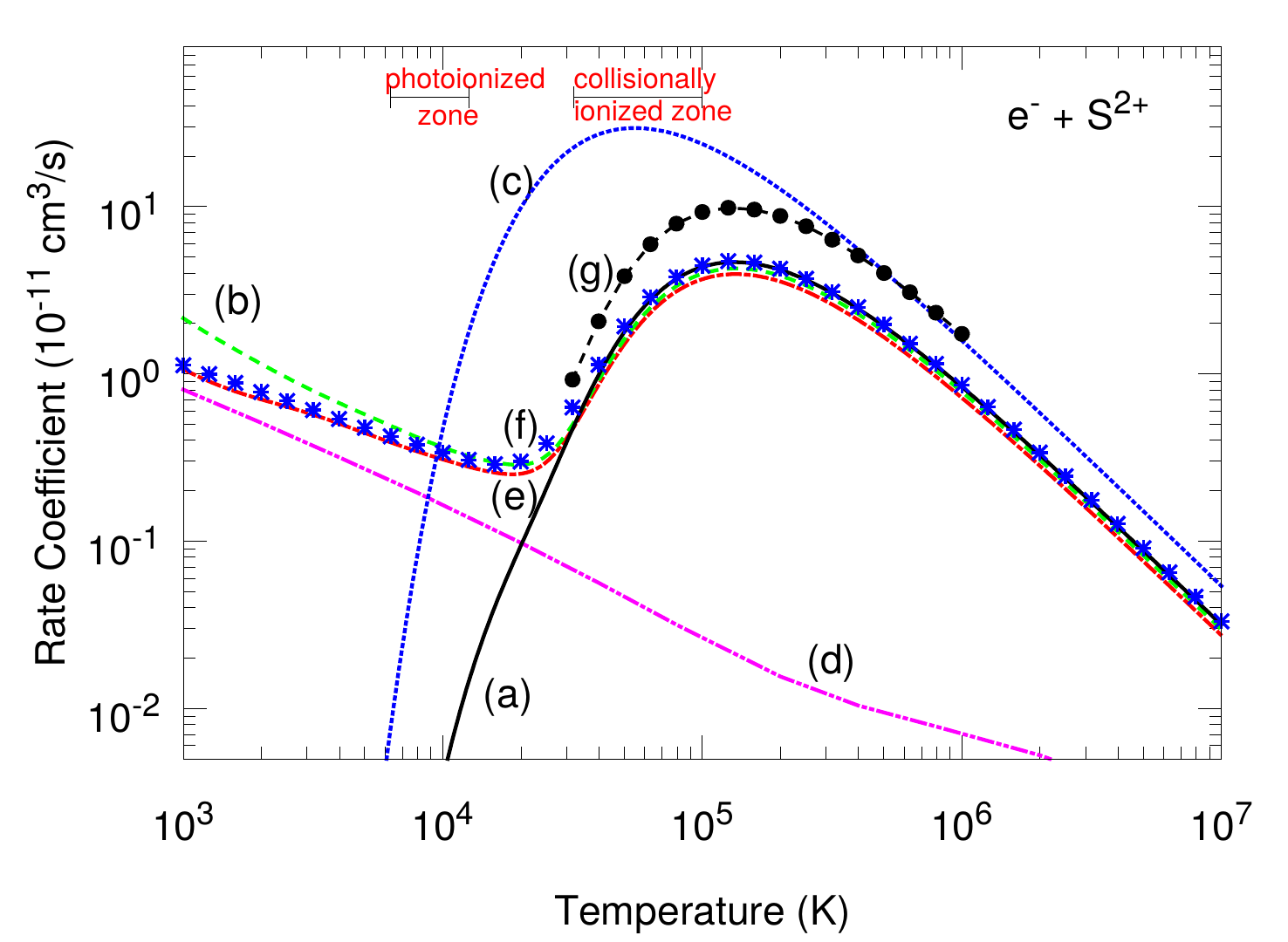}
\caption[]{\label{Fig:S}
Total Maxwellian-averaged DR and RR ground-state rate coefficients for $\rm{S^{2+}}$:
(a) \emph{black solid curve}, previous recommended data of \cite{Mazzotta:1998}; 
(b) \emph{green dashed curve}, present MCBP results; 
(c) \emph{blue dotted curve}, empirical formula of \cite{Mewe:1980};
(d) \emph{magenta dotted-dashed curve}, present RR rate coefficient;
(e) \emph{red dotted-dashed curve}, previous MCBP results \citep{Badnell:2015};
(f) \emph{blue asterisks}, $LS$ $R$-matrix, RR + DR \citep{Nahar:1995};
(g) \emph{black dashed curve with points}, $LS$ MCBP results \citep{Badnell:1991}.
}
\end{figure*}

In Fig.~\ref{Fig:S}, we show the total DR rate coefficients for the ground state of $\rm{S^{2+}}$.
A comprehensive treatment of 
$\rm{S^{2+}}$ 
DR for $3 \rightarrow 3 $ ($\Delta n_c=0$) 
core excitation has recently been performed by \cite{Badnell:2015}.
The present calculations are performed using the same MCBP IPIRDW approach
as in the previous work, but we also include
the small contributions from the $3 \rightarrow 4 $ ($\Delta n_c=1$) core excitations, 
unlike in the previous work. 
As detailed more fully in the earlier work
by \cite{Badnell:2015}, the DR resonance contributions to the rate coefficient can be classified into one of three categories.  
First, there are contributions from the well-known dipole resonances  
\citep{Burgess:1964} --- those that accumulate to a 
dipole-allowed, core-excited S$^{2+}$ thresholds, such as 
the $3s^23p3d(^3D)\, nl$ Rydberg series.  
These give rise to the characteristic high-temperature 
DR rate coefficient peak at temperatures $\frac{3}{2}kT\sim \mbox{ionization limit}$. 

As seen in Fig.~\ref{Fig:S}, both MCBP calculations are in good agreement at high temperatures.

The second category of resonance contributions is due to the fine-structure induced, core-excited states, 
such as the $3s^23p^2(^3P_{1,2})\, nl$ Rydberg series.  These resonances only contribute at low energies 
-- below the fine-structure splitting of the Si-like $3s^23p^2$ $^3P$ ground term --- and therefore the corresponding 
Rydberg series consist of high-$n$ resonances ($22\le n \le 32$ for all sequences).  
This latter series yields a DR rate coefficient that peaks (and dominates) at low temperatures 
$\frac{3}{2}kT\sim \mbox{fine-structure splitting}$.

The third category of resonances encountered are the so-called ``$(N+1)$-electron resonances'' --- low-lying $n=3$ dipole resonances 
such as $3s3p^33d$, in the case of this sequence.  
These DR resonance contributions are the most uncertain due to the corresponding uncertainty in energy position, as 
discussed in the earlier case of  S$^{2+}$ \citep{Badnell:2015}.  
To obtain total DR rate coefficients at $10^4$ K consistent with that required to determine the sulphur abundance in the 
Orion Nebula, a photoionized plasma, \cite{Badnell:2015} shifted the positions of these $n=3$ resonances to lower energies. 
This adjustment was further justified by simpler MCHF structure comparisons for the near-threshold, bound $(N+1)$-electron states 
of S$^+$, indicating that the computed $(N+1)$-electron energy positions were indeed higher than the experimental values for bound states 
\citep{Badnell:2015}. For consistency, we make the same shift for S$^{2+}$ of $\Delta E_{N+1}=-0.157$ Ryd.

Also in Fig.~\ref{Fig:S}, 
the present results are compared with other available data including the 
$LS$ coupling results of \cite{Badnell:1991}, using AUTOSTRUCTURE, and the $LS$ coupling $R$-matrix results of \cite{Nahar:1995}, 
which include both RR and DR contributions.
We note that the recommended data set of \cite{Mazzotta:1998} for $\rm{S^{2+}}$ appears to use the high-temperature  $R$-matrix results
of \cite{Nahar:1995}.

At this point, it is worth discussing the expected accuracy of our computed DR results, especially as it pertains to the three 
different categories of resonances.   The first dipole core series, as treated in the original \citet{Burgess:1964} formulation, 
peaks at a high temperature  given by the Rydberg series limit $n\rightarrow \infty$ energy positions and core oscillator strengths, 
the latter being computed fairly accurately in general.
Provided that we perform the empirical shift of each core Rydberg limit to the experimental values \citep{NIST}, 
thereby shifting every Rydberg member by the same energy, we expect to minimize the uncertainty in the high-temperature 
dipole-dominated DR rate coefficient.

In the same manner, the fine-structure resonances, that contribute strongly at lower temperatures, are governed by $n\rightarrow \infty$ Rydberg
series, with limits given by the fine-structure splitting of the S$^{2+}$ ground state.  In fact, the minimum $n$ of each Rydberg series is given 
by $-Z^2/n^2 \approx \mbox{fine-structure splitting}$, giving $n \ge 22$ for all series, and minimizing the resonance energy uncertainty. 
Provided that the calculations reproduce, or empirically shift to, the fine-structure split Si-like experimental energies \citep{NIST}, 
we minimize the uncertainty in these resonance contributions.

The third type of $(N+1)$-electron resonance contributions carry the largest uncertainty, as discussed more fully in \cite{Badnell:2015} 
for the case of S$^{2+}$.  This uncertainty in rate coefficient contribution is due to the corresponding, relatively large, 
uncertainty in the resonance energy positions of the low-lying ($n=3$) Rydberg members.  However, as we discuss further below, the $(N+1)$-electron 
states all eventually become bound for higher ionization states: only for lower charge states are some of the $(N+1)$-electron states autoionizing,
thereby contributing to DR.  Furthermore, the total uncertainties become negligible at even lower charge-states, as we will now demonstrate 
by looking at the next highest charge states: Cl$^{3+}$ and Ar$^{4+}$.

\begin{figure*}[!hbtp]
\centering
\includegraphics[angle=0,scale=1.0]{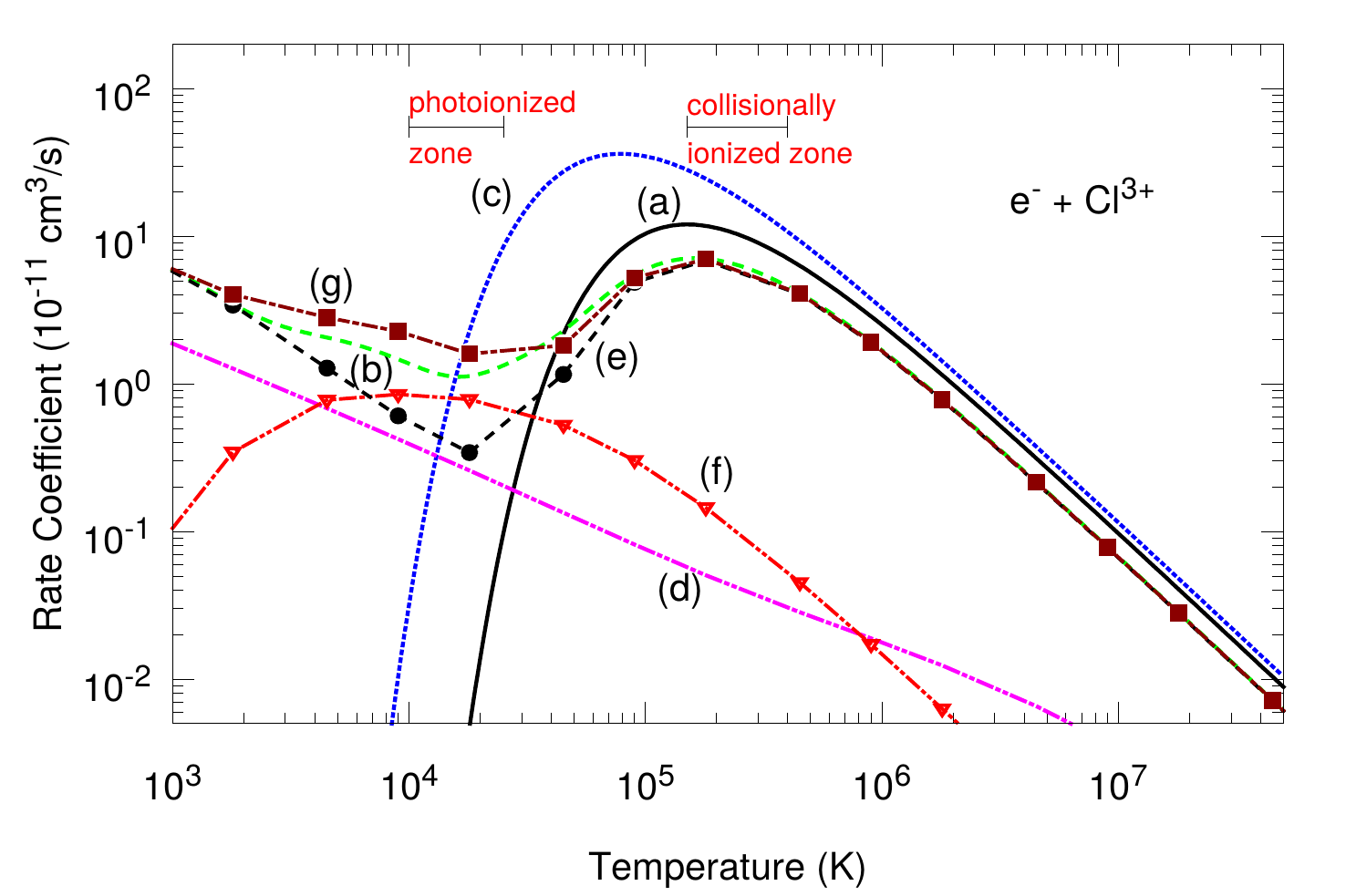}
\caption[]{\label{Fig:Cl}
Total Maxwellian-averaged DR and RR ground-state rate coefficients for $\rm{Cl^{3+}}$:
(a) \emph{black solid curve}, previous recommended data of \cite{Mazzotta:1998}; 
(b) \emph{green dashed curve}, present MCBP results; 
(c) \emph{blue dotted curve}, empirical formula of \cite{Mewe:1980};
(d) \emph{magenta dotted-dashed curve}, present RR rate coefficient;
(e) \emph{black dashed curve with filled circles}, present DR rate coefficients omitting $n=3$ resonance contributions;
(f) \emph{red dotted-dashed curve with filled triangles}, present DR rate coefficients including only the $n=3$ resonance contributions;
(g) \emph{brown dashed curve with filled squares}, present DR rate coefficients with the n=3 resonances lowered by $-0.12$~Ryd.
}
\end{figure*}

 
In Fig.~\ref{Fig:Cl}, we show total DR rate coefficients for the ground state of $\rm{Cl^{3+}}$, separating the $n=3$ $(N+1)$-electron 
contribution (about 50\% of the total for $T=2\times 10^4$K) from the total.  In view of this strong contribution and the  
uncertainties known to be associated with resonance energy uncertainties here, it is important to try and establish the temperatures
where the DR rate coefficients are affected, and by how much. Unfortunately, NIST does not give any autoionizing energies for $\rm{Cl^{2+}}$.
Furthermore, they give no bound energies for either of the two lowest lying configurations which give rise to autoionizing states
($3s^2 3p 3d^2$ and $3s 3p^3 3d$). Consequently, we can only use observed energies from $3s^2 3p^2 3d$ to guide us to a plausible
shift of the resonances. Even here the doublets and quartets show different levels of agreement and it is only practical to use a
single global shift which applies to all resonances. We choose it to be the largest difference, $\sim 0.12$ Ryd, which is
already smaller than the $\sim 0.2-0.4$~Ryd case of $\rm{S^{+}}$\citep{Badnell:2015}. 
Thus, we empirically lower the $(N+1)$-electron resonances by $-0.12$~Ryd and this gives rise to an increase the total DR rate coefficient of
$\sim 10\%$ at photoionized plasma temperatures (see Fig.~\ref{Fig:Cl}).
Our present results are also compared to the earlier results
of \cite{Mazzotta:1998} and \cite{Mewe:1980}, both of which are
based on $LS$ calculations that lack any low-temperature fine-structure DR 
contributions that are included in our calculations. For comparison, the total RR rate coefficient for $\rm{Cl^{3+}}$ is also shown.

\begin{figure*}[!hbtp]
\centering
\includegraphics[angle=0,scale=1.0]{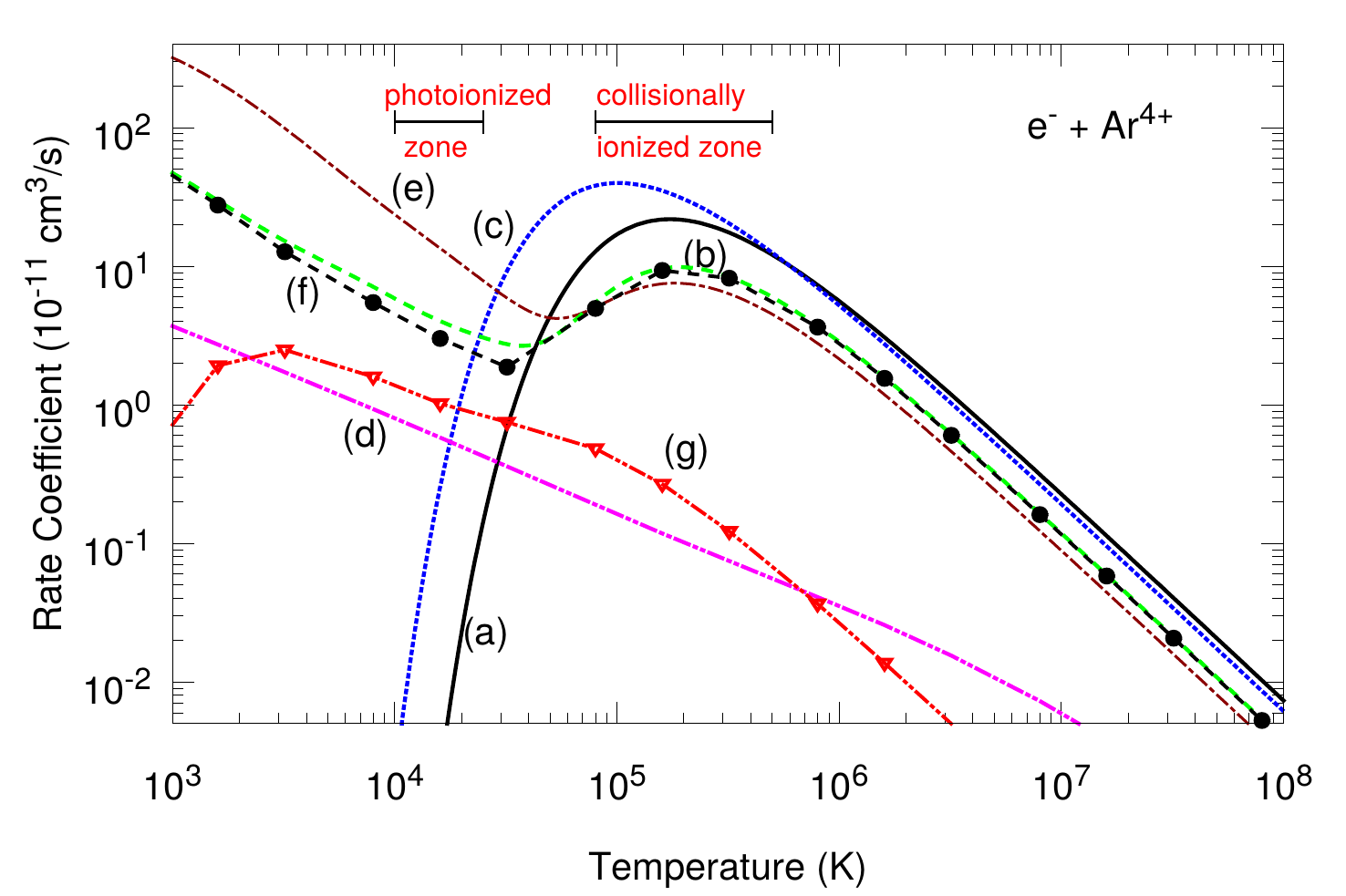}
\caption[]{\label{Fig:Ar}
Total Maxwellian-averaged DR and RR ground-state rate coefficients for $\rm{Ar^{4+}}$:
(a) \emph{black solid curve}, previous recommended data of \cite{Mazzotta:1998}; 
(b) \emph{green dashed curve}, present MCBP results; 
(c) \emph{blue dotted curve}, empirical formula of \cite{Mewe:1980};
(d) \emph{magenta dotted-dashed curve}, present RR rate coefficient;
(e) \emph{red dotted-dashed curve}, previous MCBP results of \cite{Arnold:2015}.
(f) \emph{black dashed curve with filled circles}, present DR rate coefficients omitting $n=3$ resonance contributions;
(g) \emph{red dotted-dashed curve with filled triangles}, present DR rate coefficients including only the $n=3$ resonance contributions.
}
\end{figure*}

In Fig.~\ref{Fig:Ar}, we show total DR rate coefficients for the ground state of $\rm{Ar^{4+}}$, 
again showing just the contribution from the $n=3$ resonances as well. 
For this higher-ionized system, the $3s3p^33d (^4D_{7/2})$ state that dominated the low-$T$ $\rm{S^{2+}}$ DR rate coefficient, 
due to its large oscillator strength and near-threshold positioning, is now bound.     
However, other $n=3$ resonances still contribute to the low-temperature DR --- the ionization stage is still relatively low --- 
but their contribution to the total is only about 15\% at $T=3\times 10^4$~K. Furthermore, the rate coefficient is found to be 
fairly insensitive to the uncertainty in $n=3$ resonance positions. Lowering them by $-0.12$~Ryd only results in an
increase of a few percent at photoionized plasma temperatures (too small to be shown separately).

We also compare our results to the previously published final-state level-resolved rate coefficients of \cite{Arnold:2015}. 
Although both present and previous calculations used the same methodology, there are some important differences. 
First, a different basis set of $N$-electron target configurations was used in the previous work. 
Second, the previous work also used a different scaling parameter for each orbital whereas the same scaling parameter 
was used for all the orbitals in the present work, as listed in Table~\ref{table:lamda}. Third, the previous work also 
shifted the $N$-electron target energies relative to the NIST values. We see from Fig.~\ref{Fig:Ar} that the present DR rate coefficient
is less than the previous value by about $50\%$ in the photoionized plasma zone and by about $10\%$ in the collisionally-ionized 
plasma zone. 
The low temperature difference is a little large, even allowing for its uncertainty in a low-charge ion.
We checked that the use of further observed energies had negligible effect. Instead, it appears  (Loch, private communication, 
2017) that an incorrect input dataset was used by \cite{Arnold:2015}. The intended dataset gives results much more in line
with ours. 

As seen in Fig.~ \ref{Fig:Ar}, the recommended data of \cite{Mewe:1980} and \cite{Mazzotta:1998}, based on $LS$ 
high-temperature calculations, do not reproduce the fine-structure resolved 
IC DR, for two reasons. First, the fine-structure splitting gives rise to additional Rydberg series near threshold,
thereby increasing the low-temperature DR rate coefficient. Second, as discussed in \cite{Shahin:2012}, 
at higher temperatures and for states of  sufficiently high $n$, fine-structure autoionization within terms of 
doubly-excited states, and subsequent fine-structure autoionization following radiative decay, is responsible 
for additional DR suppression, giving high-temperature IC results that are lower than the $LS$ ones.

Continuing along the series for higher ionization stages, we note that our findings for  $\rm{K^{5+}}$ are similar to those for 
$\rm{Ar^{4+}}$ --- about 15\% contribution from the $(N+1)$-electron resonances ---
while for $\rm{Ca^{6+}}$ these resonances contribute at most 5\% to the total DR rate coefficient. Higher-charged ions
have negligible contribution. Conversely, the $(N+1)$-electron resonance contributions for $\rm{P^{+}}$ are small as well, 
about 5\%, because the strongest of these resonances are high enough in energy (and remain so under any reasonable shift)
that they are dominated by, and masked by, the stronger dipole resonances. Thus, likely only for S$^{2+}$ \citep{Badnell:2015}
do we have a significant uncertainty in the total DR rate coefficients at photoionized plasma temperatures due to the
uncertainty in energy positions of the $(N+1)$-electron resonances.

\begin{figure*}[!hbtp]
\centering
\includegraphics[angle=0,scale=1.0]{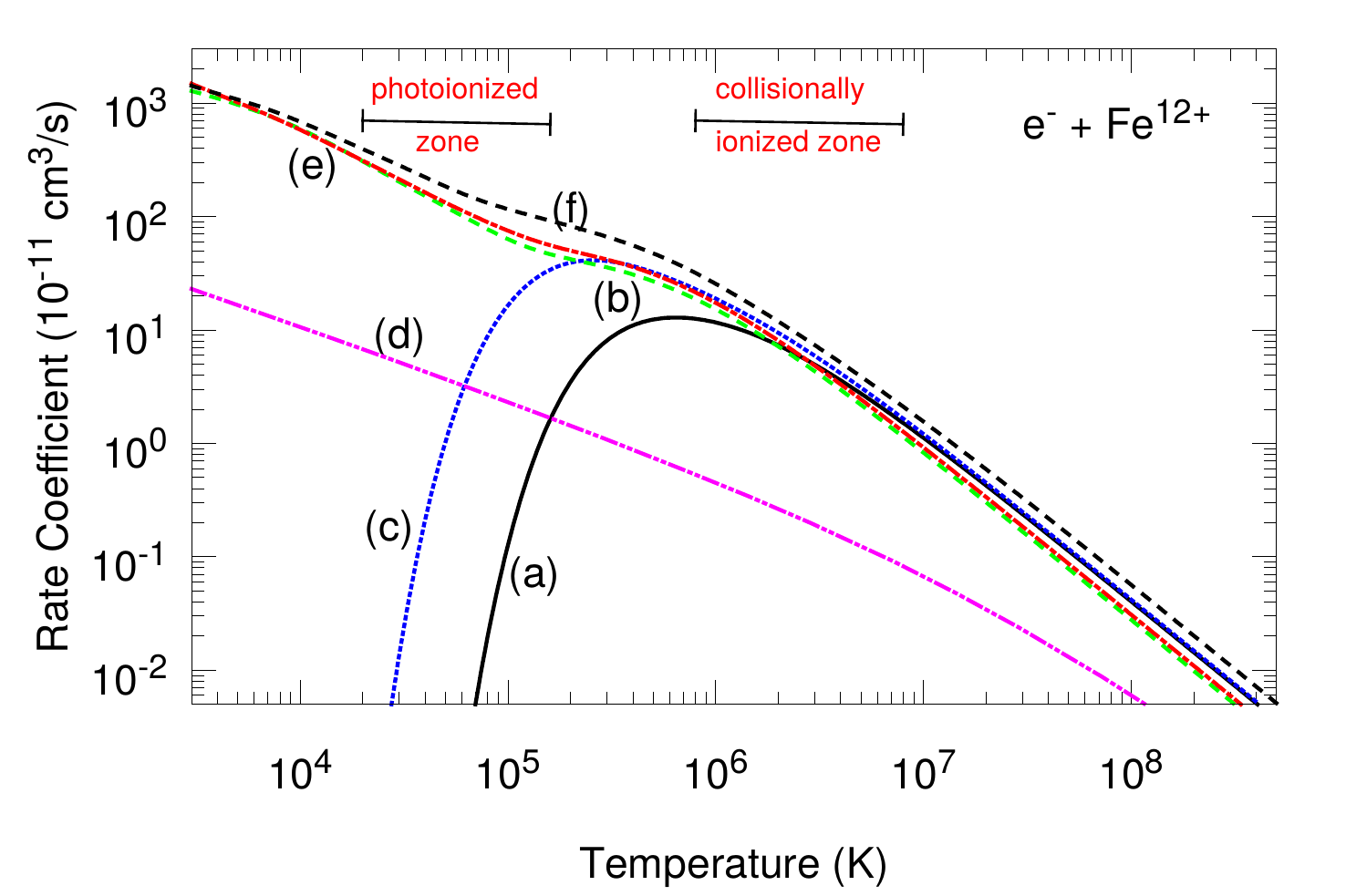}
\caption[]{\label{Fig:Fe}
Total Maxwellian-averaged DR and RR ground-state rate coefficients for $\rm{Fe^{12+}}$:
(a) \emph{black solid curve}, previous recommended data of \cite{Mazzotta:1998}; 
(b) \emph{green dashed curve}, present MCBP results; 
(c) \emph{blue dotted curve}, empirical formula of \cite{Mewe:1980};
(d) \emph{magenta dotted-dashed curve}, present RR rate coefficient;
(e) \emph{red dotted-dashed curve}, previous MCBP results presented in \cite{Hahn:2014}; and
(f) \emph{black dashed curve}, experimental measurements \citep{Hahn:2014}.
}
\end{figure*}

In Fig.~\ref{Fig:Fe}, we compare our intermediate-coupling DR rate coefficients, 
for $\rm{Fe^{12+}}$ forming $\rm{Fe^{11+}}$ via $3 \rightarrow 3$ ($\Delta n_c=0$) and $3 \rightarrow 4 $ ($\Delta n_c=1$) 
core excitations, to experimental measurements, carried out using the heavy-ion Test Storage Ring (TSR) at the 
Max-Planck Institute for  Nuclear Physics in Heidelberg \citep{Hahn:2014}. We also compare our present DR data to
previous MCBP (AUTOSTRUCTURE) calculations by \citet{Hahn:2014}, finding a difference of less than $10\%$ in the photoionized plasma region and by 
about $5\%$ in the collisionally-ionized plasma region. 
Similar to \citet{Hahn:2014}, the present theoretical rate coefficient is smaller than the experimental value, 
by approximately $30\%$ in the photoionized region and $25\%$ in the collisionally ionized region. 
This somewhat largish discrepancy cannot be explained by any inaccuracies discussed earlier for the three different types of resonances 
(indeed, all $(N+1)$-electron states are strongly bound by Fe$^{12+}$). Instead, a fourth type of resonance contribution error was discussed  
by \cite{Hahn:2014}.  At higher ionization stages, the core-excited $N$-electron states, such as $3s^23pn_cl_c$ ($3<n_c<\infty$) contribute more 
to the total DR, and our computational termination at $n_c=4$ means that $5\le n_c<\infty$ contributions are neglected.  
These could account for much of the discrepancy: assuming a pure $n_c^{-3}$ scaling beyond $n_c=4$, 
the contributions from $n_c\ge5$ increase the
total DR rate coefficient by 15--25\% over $10^6 - 10^9$~K. But, it should be noted that increasing Auger suppression
with increasing $n_c$ can be expected to reduce this amount somewhat.
On the other hand, it cannot be ruled out that the experiment 
is in fact too high by about 25\%, which is roughly the total calibration uncertainty in the experiments.
Finally, we note also that the previous results of \citet{Mazzotta:1998,Mewe:1980}, that are based on $LS$-coupling
calculations, do not take into account fine-structure-induced DR and therefore do not show any low-temperature enhancement, 
as seen in Fig.~\ref{Fig:Fe}.

\begin{figure*}[!hbtp]
\centering
\includegraphics[angle=0,scale=1.0]{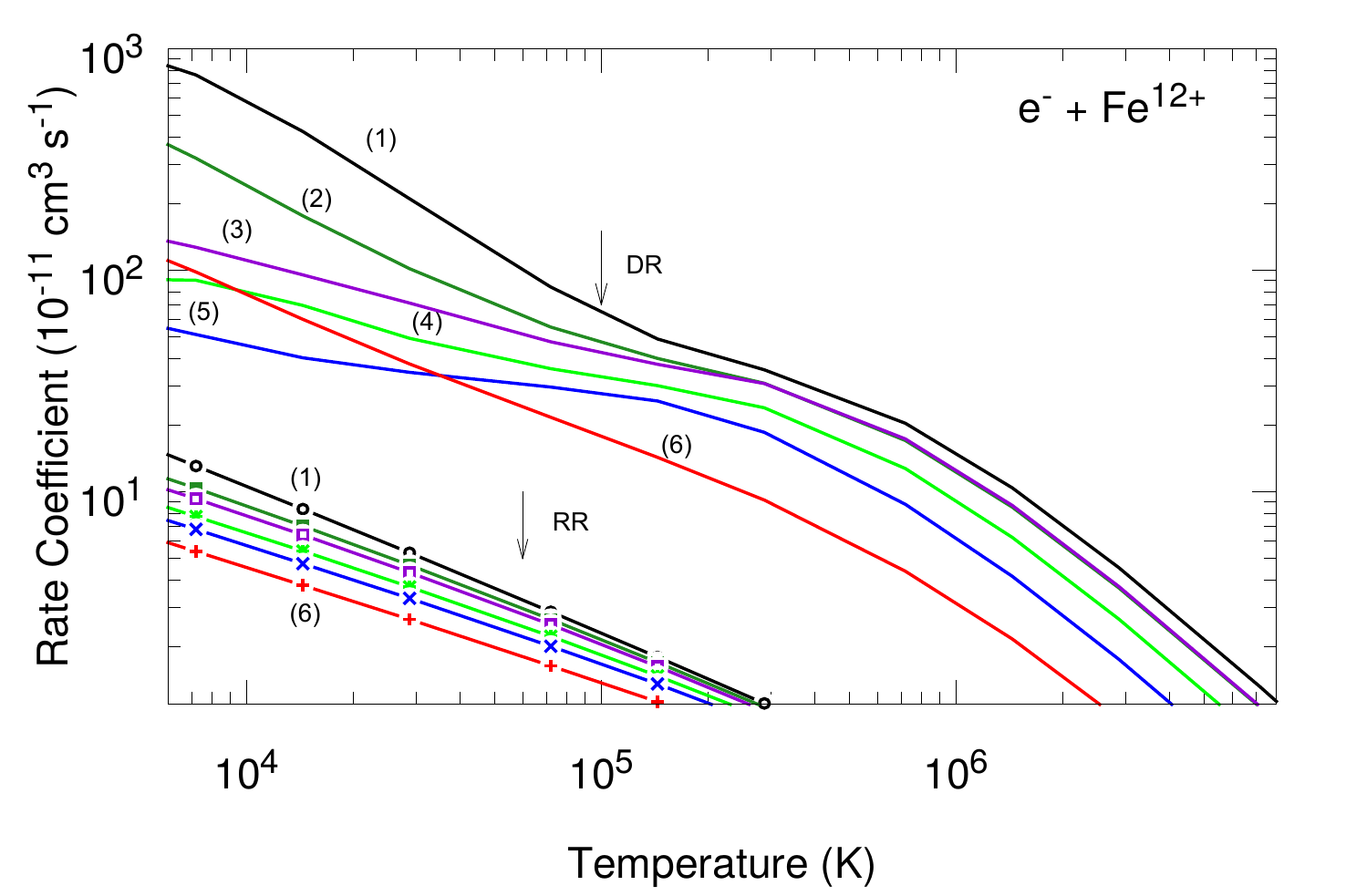}
\caption[]{\label{Fig:metastables}
Total Maxwellian-averaged DR and RR rate coefficients from the ground ($3s^23p^2~(^{3}P_0)$ ($i=1$)) and metastable initial levels ($3s^23p^2~(^3P_{1,2},^1D_{2},^1S_{0})$ ($i=2-5$) and $3s^23p3d~(^5S_2)$ ($i=6$)) of $\rm{Fe^{12+}}$.
}
\end{figure*}

In Fig.~\ref{Fig:metastables}, we show the total DR rate coefficients from metastable  as well as ground initial states 
of $\rm{Fe^{12+}}$. We note first that the DR rate coefficients are $LS$-term dependent. Second, at low temperature
there is a significant difference among DR from the three fine-structure split levels $3s^23p^2~(^3P_{0})$, 
$3s^23p^2~(^3P_{1})$ and $3s^23p^2~(^3P_{2})$. The $3s^23p^2~(^3P_{0})$ level-resolved DR is enhanced by second and third 
fine-structure-split Rydberg series near threshold whereas the $3s^23p^2~(^3P_{1})$ DR has only the 
second fine-structure-split Rydberg series near threshold, and the $3s^23p^2~(^3P_{2})$ series has no fine-structure-split 
Rydberg series enhancement near threshold. Also shown in Fig.~\ref{Fig:metastables} are the present RR results from both the ground 
and metastable initial levels  of $\rm{Fe^{12+}}$.

\begin{figure*}[!hbtp]
\centering
\subfigure{\resizebox{0.48\textwidth}{!}{\includegraphics[angle=0,scale=1.0]{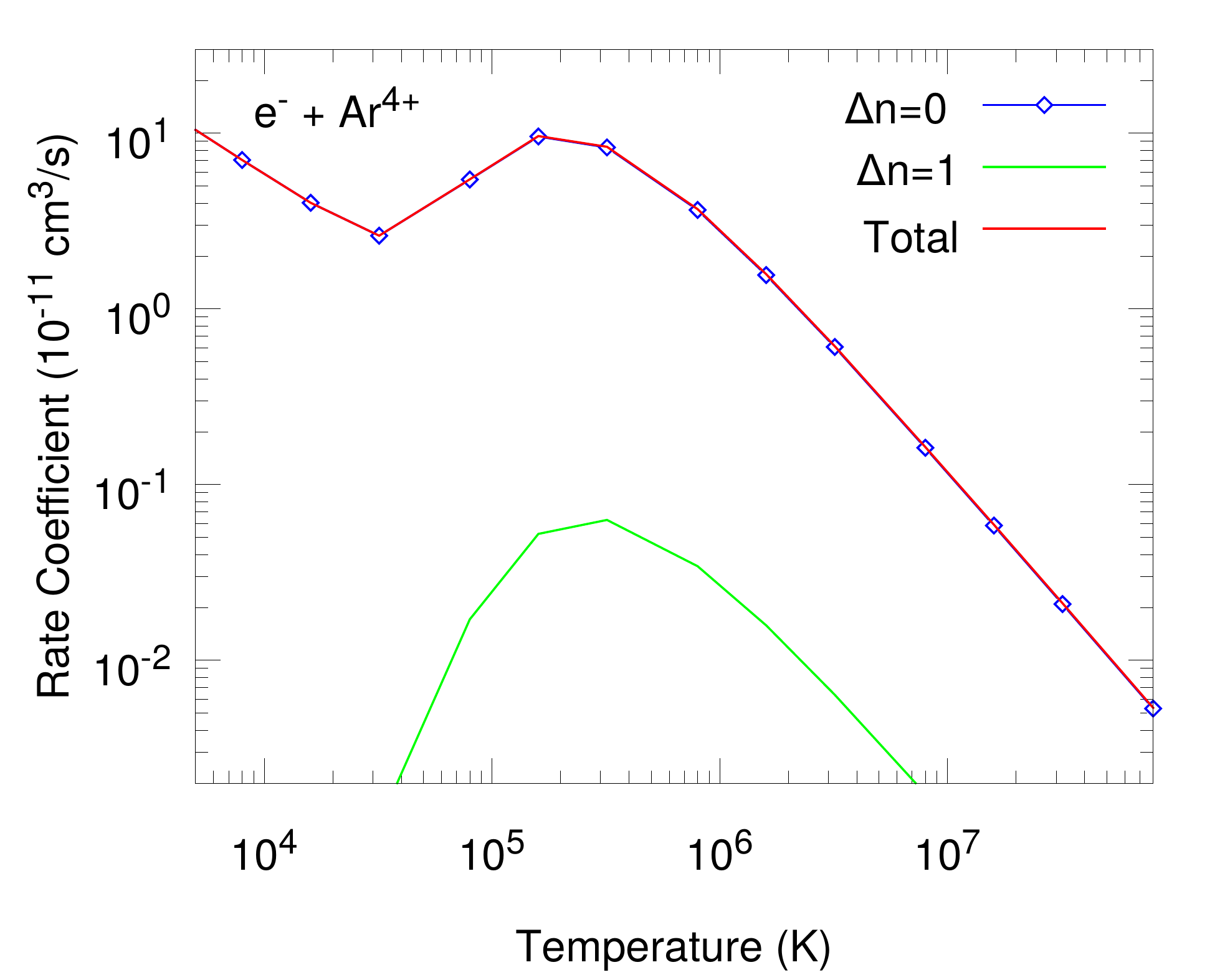} }\label{Fig:Ar_n}}
\subfigure{\resizebox{0.48\textwidth}{!}{\includegraphics[angle=0,scale=1.0]{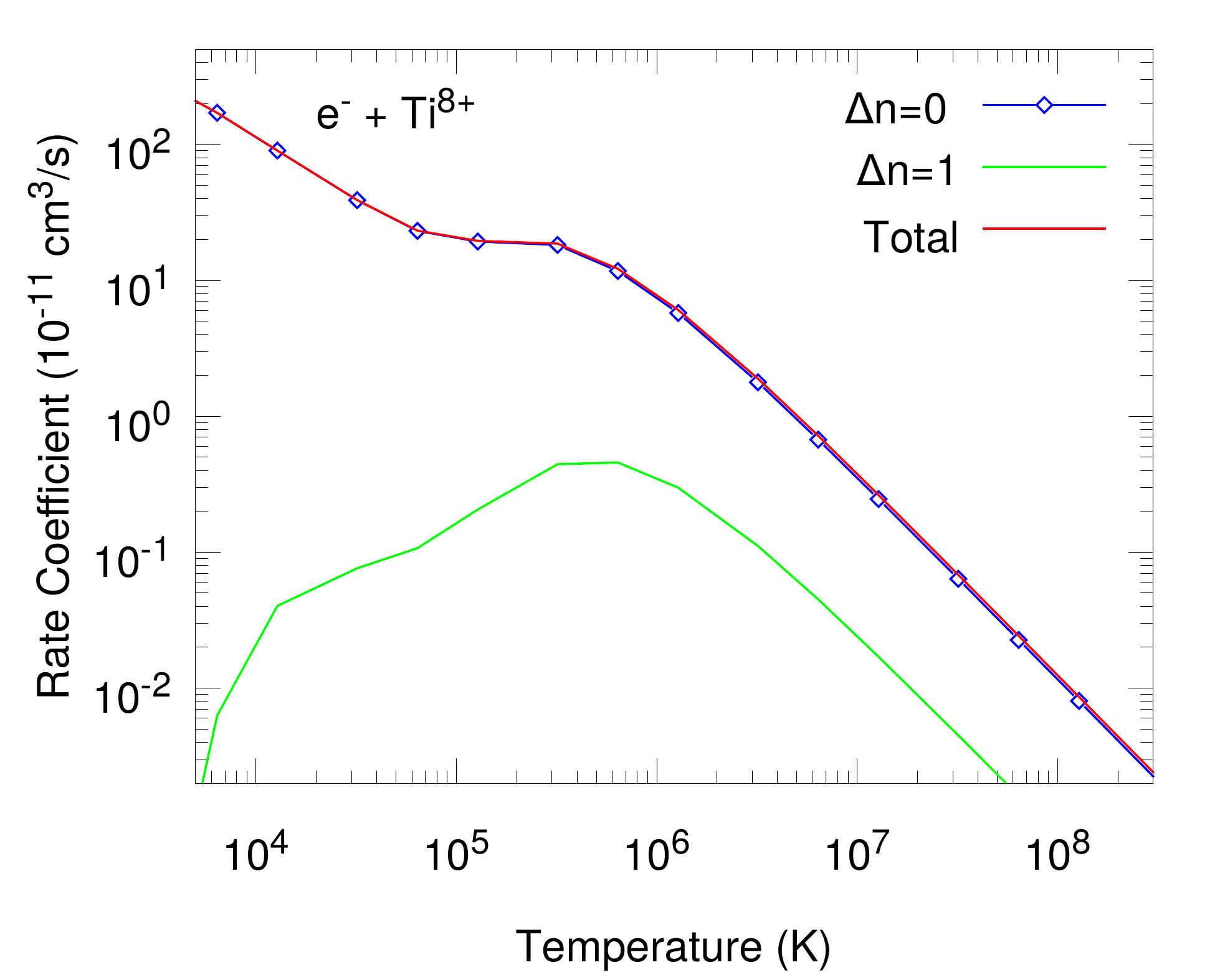} }\label{Fig:Ti_n}}
\subfigure{\resizebox{0.48\textwidth}{!}{\includegraphics[angle=0,scale=1.0]{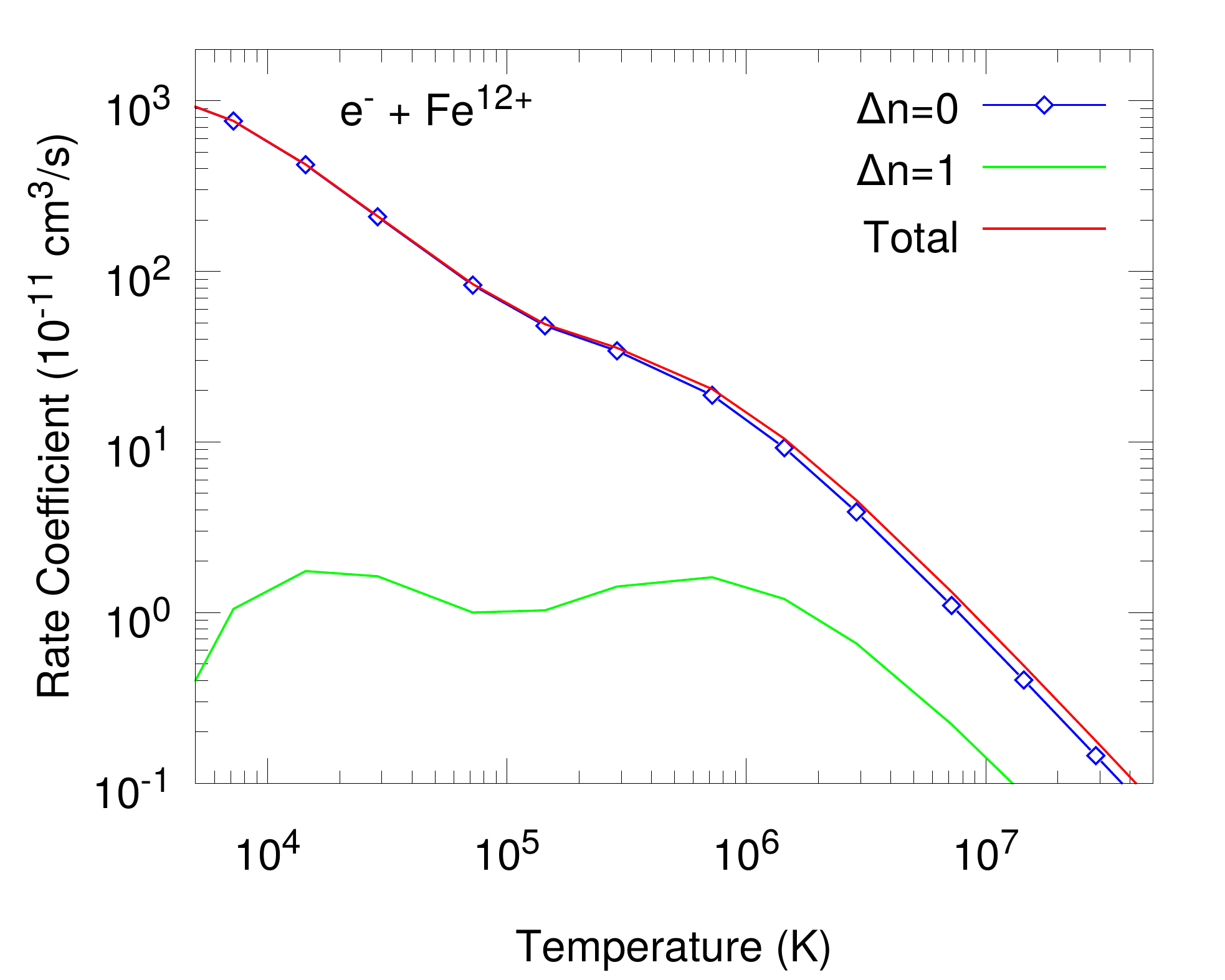} }\label{Fig:Fe_n}}
\subfigure{\resizebox{0.48\textwidth}{!}{\includegraphics[angle=0,scale=1.0]{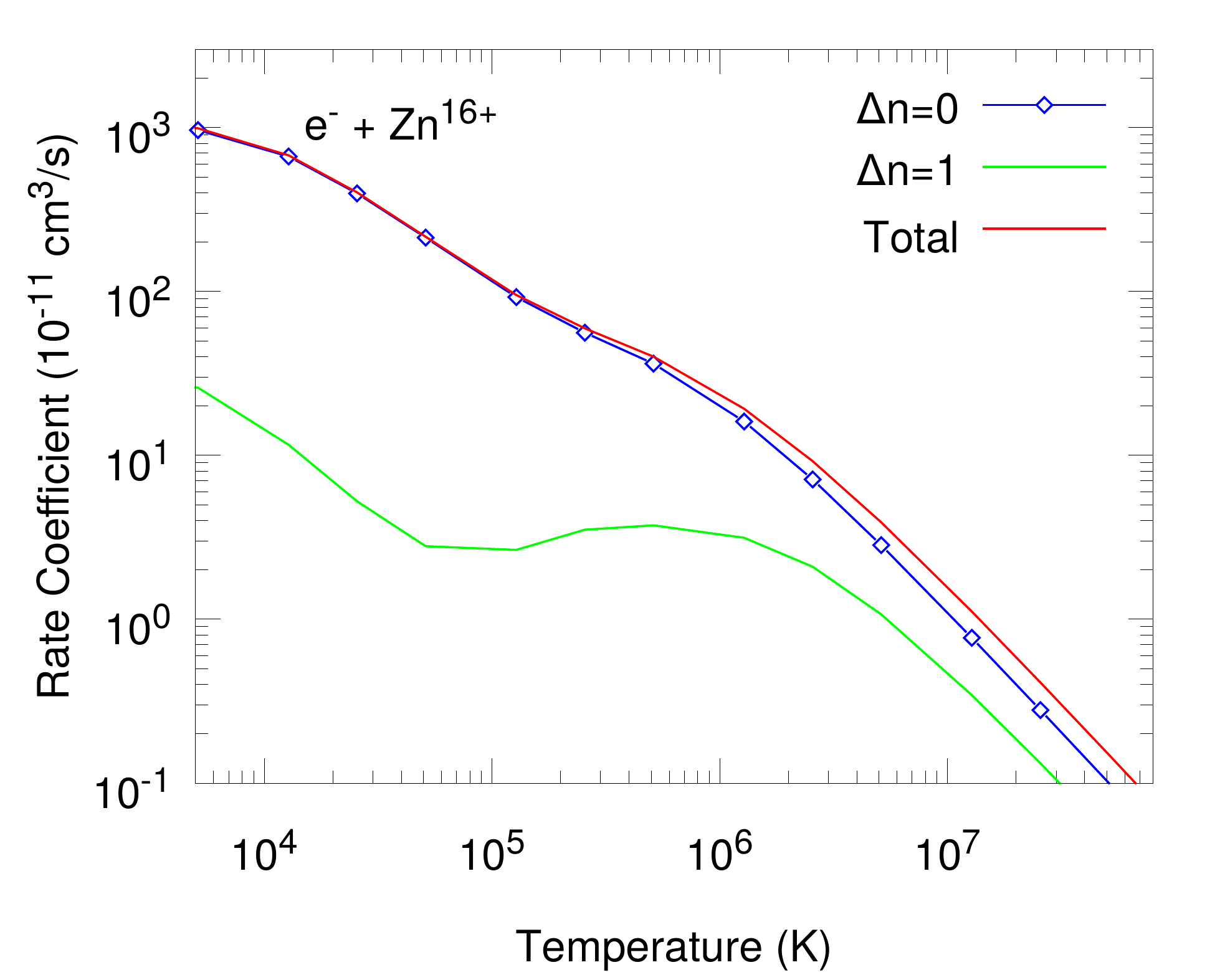} }\label{Fig:Zn_n}}
\caption[]{\label{Fig:Ar_n:Zn_n}
Total Maxwellian-averaged ground-level DR rate coefficients and the separate contributions from
$\Delta n_c=0$ and $\Delta n_c=1$
core excitations for $\rm{Ar^{4+}}$, $\rm{Ti^{8+}}$, $\rm{Fe^{12+}}$, and $\rm{Zn^{16+}}$ ions. 
}
\end{figure*}

We present in Fig.~\ref{Fig:Ar_n:Zn_n} the DR rate coefficients for both 
$\Delta n_c = 0$ ($3 \rightarrow 3$) and $\Delta n_c = 1$ ($3 \rightarrow 4$) core excitations, for selected ions 
along the silicon-like sequence. Also shown are the total ($\Delta n_c = 0$ + $\Delta n_c = 1$) DR rate coefficients. 
For low-charged $\rm{Ar^{4+}}$, the contribution from $3 \rightarrow 4$ core excitation to the total DR rate 
coefficient is negligible. Additionally, using a configuration-averaged distorted wave method, \cite{Arnold:2015} 
also showed that the contribution from $\Delta n_c = 2$ core excitation to the total rate coefficient is $3$ to $4$ 
orders of magnitude smaller than the sum of the contributions from $\Delta n_c = 0$ and $\Delta n_c = 1$ core excitations. 
The $3 \rightarrow 4$ core excitation contributions are less than  $1\%$ for $\rm{Ti^{8+}}$, and $2\%$ for $\rm{V^{9+}}$, 
whereas, for $\rm{Zn^{16+}}$, the $\Delta n_c = 1$ contribution leads to an increase of approximately $15\%$ in the total 
DR rate coefficient.

\begin{figure*}[!hbtp]
\centering
\includegraphics[angle=0,scale=1.0]{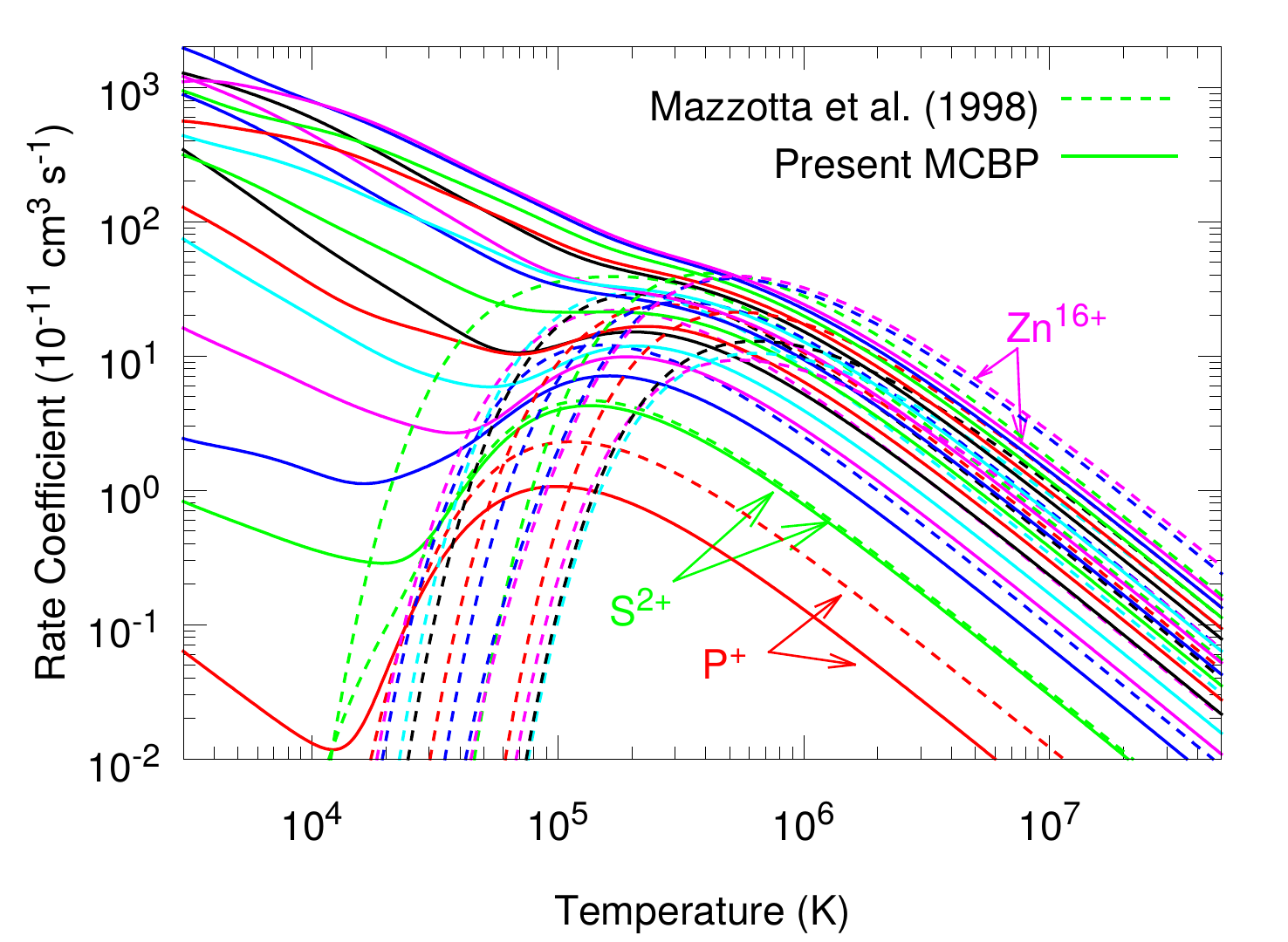}
\caption[]{\label{Fig:Mazzotta}
Comparison between the present total Maxwellian-averaged IC DR rate coefficients (solid curves) and the recommended data 
\cite[dashed curves]{Mazzotta:1998} for the silicon-like isoelectronic sequence.
}
\end{figure*}

Lastly, in Fig.~\ref{Fig:Mazzotta}, we compare our present Maxwellian-averaged DR rate coefficients (in IC), 
for the entire silicon-like isoelectronic 
sequence, to the recommended data of \cite{Mazzotta:1998}. The recommended data is based-upon
calculations that do not take explicit account of fine-structure and therefore do not show DR contributions
at low temperatures that arise from
fine-structure Rydberg series near  threshold. This deficiency becomes greater with increase in the effective charge 
$z$. At higher temperatures, the two sets of results differ appreciably, especially for low-z ions. For example, there is a difference 
of about $50\%$ for {\bf $\rm{P^{+}}$} and $30\%$ for {\bf $\rm{Cl^{3+}}$}.

Also, note that the final results and the fitting coefficients listed in Table \ref{table:dr_all} correspond to the unshifted calculations.


\section{Summary}
\label{Sec:Summary}
We have carried-out multi-configuration intermediate-coupling Breit-Pauli calculations for total and partial 
(final-state level-resolved) DR and RR rate coefficients for all ions from $\rm{P^{+}}$ through $\rm{Zn^{16+}}$ 
of the silicon-like isoelectronic sequence. 
We have compared total dielectronic recombination rate coefficients with other theoretical and experimental 
results. Good agreement is found at higher temperatures. At lower temperatures that are applicable to photoionized 
plasmas, our new results include additional DR contributions that were not included in most previous 
results, and  differ markedly from the
recommended results of \cite{Mazzotta:1998}. 
We have also investigated the contributions from the low-lying $(N+1)$-electron resonances to the 
low-temperature total DR rate coefficient.  
The uncertainties associated with these contributions are likely significant only for S$^{2+}$ \citep{Badnell:2015}.
Fitting coefficients for total DR and RR rate coefficients from the ground state were presented.
Partial DR and RR rate 
coefficients are archived in OPEN-ADAS\footnote{http://open.adas.ac.uk} using the ADAS \emph{adf09} and 
\emph{adf48} formats, respectively. 
These data are needed for both astrophysical and fusion plasma modelling and constitute 
part of a dielectronic recombination database assembly for  
modelling dynamic finte-density plasmas in general \citep{Badnell:2003}. 





%
%
%
%
%
%
%
%
%
%
%
%
\begin{table*}[!htbp]
\caption{\label{table:lamda}
Radial scaling parameters for orbitals ($1s$, $2s$, $2p$, $3s$, $3p$, $3d$, $4s$, $4p$, $4d$ and $4f$) used in the 
present study for $\Delta n_c= 0$  and $\Delta n_c= 1$
core-excitations in the silicon-like isoelectronic sequence.}
\centering \small 
\begin{tabular}{c c c c c c c c c c c}
\\[0.1in]
\hline
\hline
  Ion  & $\rm{P^{+}}$  & $\rm{S^{2+}}$ & $\rm{Cl^{3+}}$ & $\rm{Ar^{4+}}$ & 
$\rm{K^{5+}}$ & $\rm{Ca^{6+}}$ & $\rm{Sc^{7+}}$ & $\rm{Ti^{8+}}$ & $\rm{V^{9+}}$ & 
$\rm{Cr^{10+}}$~--~$\rm{Zn^{16+}}$ \\
\hline
\multicolumn{11}{c}{\vspace*{-2mm}} \\
$\lambda_{nl}$   &  1.114 & 1.13 & 1.12 & 1.125 & 1.15 & 1.15 & 1.16 & 1.17 & 1.18 & 1.185   \\
\multicolumn{11}{c}{\vspace*{-2mm}} \\
\hline
\hline
\end{tabular}
\end{table*}
%
%
%
\begin{table*}[!htbp]
\caption{\label{table:dr_all} Fitting coefficients
$c_i$ (in cm$^3$ K$^{3/2}$ s$^{-1}$) and $E_i$ (in K) for the total ground-state IC DR rate coefficients.}
\begin{tabular}{l c c c c c c c c}
\hline   
\hline
ion        & $c_1$ & $c_2$      &  $c_3$     & $c_4$     & $c_5$     & $c_6$   & $c_7$ & $c_8$ \\
\hline
\multicolumn{8}{c}{\vspace*{-2mm}} \\
$\rm{P^{+}}$   & 5.128E-08 & 2.744E-08 & 2.647E-08 &  3.652E-08 & 2.286E-06 & 9.048E-04 &  5.975E-04 & 4.853E-05 \\
$\rm{S^{2+}}$  & 3.040E-07 & 4.393E-07 & 1.609E-06 & 4.980E-06 & 3.457E-05 & 8.617E-03 & 9.284E-04 & $\cdots$ \\
$\rm{Cl^{3+}}$  & 6.019E-07 & 3.244E-06 & 2.686E-05 & 8.945E-04 & 2.095E-02 & $\cdots$ & $\cdots$  & $\cdots$ \\
$\rm{Ar^{4+}}$ & 1.590E-05 & 1.636E-05 & 7.566E-05 & 3.805E-04 & 5.247E-03 & 3.272E-02 & 1.060E-04 &   $\cdots$ \\
$\rm{K^{5+}}$   &  8.624E-05 & 8.801E-05 & 1.934E-04 & 1.878E-03 & 4.936E-02 & 3.667E-03 & $\cdots$ & $\cdots$  \\
$\rm{Ca^{6+}}$ & 4.836E-04 & 3.208E-04 & 9.281E-04 & 5.307E-02 & 2.175E-02 & $\cdots$ & $\cdots$  & $\cdots$  \\
$\rm{Sc^{7+}}$  & 1.651E-04 & 2.659E-04 & 2.517E-03 & 7.847E-02 & 1.681E-02 & $\cdots$  & $\cdots$ & $\cdots$  \\
$\rm{Ti^{8+}}$  & 4.657E-04 & 1.194E-03 & 2.720E-03 & 3.812E-02 & 8.260E-02 & $\cdots$  & $\cdots$  & $\cdots$ \\
$\rm{V^{9+}}$   & 1.903E-03 & 2.573E-03 & 5.861E-03 & 5.971E-02 & 8.198E-02 & $\cdots$  & $\cdots$  & $\cdots$ \\
$\rm{Cr^{10+}}$ & 2.687E-03 & 4.814E-03 & 6.995E-03 & 9.119E-02 & 8.006E-02 & $\cdots$  & $\cdots$ & $\cdots$  \\
$\rm{Mn^{11+}}$ & 1.040E-03 & 4.548E-03 & 1.134E-02 & 1.253E-01 & 8.492E-02 & $\cdots$  & $\cdots$ & $\cdots$ \\
$\rm{Fe^{12+}}$ & 4.469E-03 & 8.538E-03 & 1.741E-02 & 1.630E-01 & 8.680E-02 & $\cdots$ & $\cdots$  & $\cdots$ \\
$\rm{Co^{13+}}$ & 3.163E-03 & 1.128E-02 & 2.548E-02 & 1.987E-01 & 9.730E-02 & $\cdots$  & $\cdots$ & $\cdots$ \\
$\rm{Ni^{14+}}$ & 3.306E-03 & 1.699E-02 & 3.525E-02 & 2.401E-01 & 1.102E-01 & $\cdots$  & $\cdots$ & $\cdots$ \\
$\rm{Cu^{15+}}$ & 7.276E-03 & 2.120E-02 & 4.385E-02 & 2.783E-01 & 1.308E-01 & 1.513E-03 & $\cdots$ & $\cdots$ \\
$\rm{Zn^{16+}}$ & 9.796E-03 & 2.209E-02 & 5.492E-02 & 3.200E-01 & 1.495E-01 & $\cdots$  & $\cdots$ & $\cdots$ \\
\multicolumn{9}{c}{\vspace*{-2mm}} \\
\hline
\hline
\multicolumn{9}{c}{\vspace*{-2mm}} \\
ion       &  $E_1$      &  $E_2$      & $E_3$       & $E_4$      & $E_5$      &   $E_6$ &   $E_7$ &   $E_8$ \\
\hline
$\rm{P^{+}}$   & 1.684E+01 & 1.053E+02 & 4.273E+02 & 7.150E+03 & 5.600E+04 & 1.399E+05 & 1.676E+05 & 3.120E+07 \\
$\rm{S^{2+}}$   & 5.016E+01 & 3.266E+02 & 3.102E+03 & 1.210E+04 & 4.969E+04 & 2.010E+05 & 2.575E+05 & $\cdots$  \\
$\rm{Cl^{3+}}$  & 1.077E+02 & 8.933E+02 & 9.908E+03 & 8.465E+04 & 2.657E+05 & $\cdots$ & $\cdots$ & $\cdots$ \\
$\rm{Ar^{4+}}$  & 2.879E+02 & 1.717E+03 & 9.917E+03 & 5.769E+04 & 2.178E+05 & 3.191E+05 & 1.250E+06 & $\cdots$\\
$\rm{K^{5+}}$   & 1.876E+02 & 2.406E+03 & 1.482E+04 & 9.215E+04 & 3.473E+05 & 4.806E+05 & $\cdots$ & $\cdots$  \\
$\rm{Ca^{6+}}$  & 3.497E+02 & 2.664E+03 & 3.433E+04 & 3.149E+05 & 6.358E+05 & $\cdots$  & $\cdots$ & $\cdots$\\
$\rm{Sc^{7+}}$  & 3.808E+02 & 4.268E+03 & 5.687E+04 & 3.742E+05 & 8.285E+05 & $\cdots$  & $\cdots$ & $\cdots$\\
$\rm{Ti^{8+}}$  & 8.056E+02 & 6.038E+03 & 3.964E+04 & 2.517E+05 & 6.006E+05 & $\cdots$  & $\cdots$ & $\cdots$\\
$\rm{V^{9+}}$   & 1.360E+03 & 7.173E+03 & 5.168E+04 & 3.081E+05 & 7.209E+05 & $\cdots$  & $\cdots$ & $\cdots$ \\
$\rm{Cr^{10+}}$ & 1.403E+03 & 8.418E+03 & 6.683E+04 & 3.746E+05 & 8.827E+05 & $\cdots$  & $\cdots$ & $\cdots$ \\
$\rm{Mn^{11+}}$ & 1.521E+03 & 1.177E+04 & 7.406E+04 & 4.268E+05 & 1.058E+06 & $\cdots$  & $\cdots$ & $\cdots$ \\
$\rm{Fe^{12+}}$ & 2.462E+03 & 1.261E+04 & 9.330E+04 & 4.887E+05 & 1.312E+06 & $\cdots$  & $\cdots$ & $\cdots$ \\
$\rm{Co^{13+}}$ & 3.776E+03 & 1.896E+04 & 1.067E+05 & 5.476E+05 & 1.588E+06 & $\cdots$  & $\cdots$ & $\cdots$ \\
$\rm{Ni^{14+}}$ & 2.329E+03 & 1.982E+04 & 1.156E+05 & 6.074E+05 & 1.911E+06 & $\cdots$  & $\cdots$ & $\cdots$ \\
$\rm{Cu^{15+}}$ & 2.466E+03 & 2.273E+04 & 1.338E+05 & 6.643E+05 & 2.220E+06 & 1.217E+07 & $\cdots$ & $\cdots$ \\
$\rm{Zn^{16+}}$ & 5.083E+03 & 2.462E+04 & 1.588E+05 & 7.384E+05 & 2.615E+06 & $\cdots$  & $\cdots$ & $\cdots$ \\
\multicolumn{9}{c}{\vspace*{-2mm}} \\
\hline
\hline
\end{tabular}
\end{table*}
\clearpage
\begin{table*}[!htbp]
\caption{ \label{table:rr_all} RR fitting coefficients for the ground states of Si-like ions (see Eq. \ref{Eq:Fit:RR}).}
\begin{tabular}{lcccccc}
\hline \hline
 ion      &     $A$    &   $B$ &  $T_{0}$ &  $T_{1}$ & $C$  &    $T_{2}$    \\
          & (cm$^3~$s$^{-1}$)&    & (K)      &  (K)     &      &   (K)        \\
\hline
\hline
$\rm{P^{+}}$    & 1.505E-09 & 0.8452 & 1.707E-02 & 1.301E+06 & 0.2467 & 1.284E+06 \\
$\rm{S^{2+}}$   & 2.478E-11 & 0.4642 & 3.294E+02 & 2.166E+07 & 0.3351 & 7.630E+05 \\
$\rm{Cl^{3+}}$  & 1.602E-10 & 0.6129 & 5.154E+01 & 3.056E+07 & 0.1342 & 6.808E+05\\
$\rm{Ar^{4+}}$  & 3.939E-10 & 0.6607 & 3.207E+01 & 3.043E+07 & 0.0761 & 6.360E+05 \\
$\rm{K^{5+}}$   & 6.034E-10 & 0.6803 & 3.503E+01 & 3.022E+07 & 0.0561 & 5.412E+05 \\
$\rm{Ca^{6+}}$  & 1.427E-08 & 0.7285 & 3.790E-01 & 3.977E+07 & $\cdots$ & $\cdots$ \\
$\rm{Sc^{7+}}$  & 9.729E-09 & 0.7294 & 1.219E+00 & 3.765E+07 & $\cdots$ & $\cdots$ \\
$\rm{Ti^{8+}}$  & 7.528E-09 & 0.7276 & 3.009E+00 & 3.764E+07 & $\cdots$ & $\cdots$ \\
$\rm{V^{9+}}$   & 8.439E-09 & 0.7289 & 3.806E+00 & 3.791E+07 & $\cdots$ & $\cdots$ \\
$\rm{Cr^{10+}}$ & 4.834E-09 & 0.7238 & 1.384E+01 & 3.952E+07 & $\cdots$ & $\cdots$ \\
$\rm{Mn^{11+}}$ & 2.598E-09 & 0.7163 & 5.391E+01 & 4.166E+07 & $\cdots$ & $\cdots$ \\
$\rm{Fe^{12+}}$ & 1.984E-09 & 0.7101 & 1.158E+02 & 4.400E+07 & $\cdots$ & $\cdots$ \\
$\rm{Co^{13+}}$ & 1.839E-09 & 0.7055 & 1.764E+02 & 4.654E+07 & $\cdots$ & $\cdots$ \\
$\rm{Ni^{14+}}$ & 1.551E-09 & 0.6991 & 3.080E+02 & 4.948E+07 & $\cdots$ & $\cdots$ \\
$\rm{Cu^{15+}}$ & 1.663E-09 & 0.6970 & 3.534E+02 & 5.214E+07 & $\cdots$ & $\cdots$ \\
$\rm{Zn^{16+}}$ & 1.506E-09 & 0.6920 & 5.288E+02 & 5.548E+07 & $\cdots$ & $\cdots$ \\
\hline
\hline
\end{tabular}
\end{table*}

\end{document}